# Are Supernovae Recorded in Indigenous Astronomical Traditions?

Duane W. Hamacher

Nura Gili Indigenous Programs Unit
University of New South Wales, Sydney, NSW, 2052, Australia
Email: d.hamacher@unsw.edu.au

**Abstract:** Novae and supernovae are rare astronomical events that would have had an influence on the sky-watching peoples who witnessed them. Although several bright novae/supernovae have been visible during recorded human history, there are many proposed but no confirmed accounts of supernovae in oral traditions or material culture. Criteria are established for confirming novae/supernovae in oral and material culture, and claims from around the world are discussed to determine if they meet these criteria. Australian Aboriginal traditions are explored for possible descriptions of novae/supernovae. Although representations of supernovae may exist in Indigenous traditions, there are currently no confirmed accounts in Indigenous oral or material traditions.

1   INTRODUCTION

Like many Indigenous cultures around the world, Aboriginal Australians possess detailed knowledge of the night sky, using it for navigation, calendars and time-keeping, food economics, ceremony, and social structure (e.g. Cairns and Harney, 2003; Fredrick, 2008; Hamacher and Norris, 2011a; Johnson, 1998). This sky knowledge involves an understanding and explanation of planetary motions, relative stellar positions, lunar phases and tides, and the position of the rising and setting sun throughout the year with respect to the landscape (Hamacher and Norris, 2011a; Norris and Hamacher, 2009; Norris et al., 2013). This knowledge includes explanations of transient phenomena, such as meteors, comets, eclipses, and aurorae (Hamacher and Norris, 2010, 2011b, 2011c; and Hamacher, 2013, respectively). This sky knowledge has been passed down through successive generations in the form of oral tradition and material culture (Clunies-Ross, 1986).

Indigenous astronomical traditions contain scientific information that to explains the natural world in terms of cause-effect. This scientific information was based on observation and deduction and was used for predictive purposes. Indigenous people linked lunar phases to tides and used this information as a guide for deciding when to fish, and their traditions contained an explanation as to how and why the moon was connected to tides (e.g. Johnson, 1998: 27, 37). The heliacal rising of the Pleiades signalled the arrival of winter in the Central Desert (Tindale, 2005: 374), while the rising of the celestial emu after dusk informed Aboriginal people that emu eggs were ready to be collected (e.g. Fuller et al., 2014 – this volume). Similarly, transient phenomenon, whether rare or common, are often linked to special events on the earth. For example, the sudden flash of a meteor may coincide with a death in the community, or the appearance of a comet or eclipse might coincide with a famine, drought, or battle (Johnson, 1998: 86-89).

## 2    NOVAE AND SUPERNOVAE

Novae are stellar explosions that occur in white dwarf stars. If a binary system consists of a white dwarf, and the companion star (generally a main sequence or red giant) is within the dwarf's Roche Lobe, the dwarf will accrete material (mostly hydrogen and helium) from the companion until it reaches a critical pressure and temperature. At this point, the process ignites a thermonuclear explosion. This event may be visible to us and can be seen with the naked eye for a period of days to weeks.

A supernova is the explosion of a star and the most violent event in the universe. The energy released during a supernova is up to 10 billion times the star's normal energy output. These events can occur in one of two primary ways: 1) A compact star, such as a white dwarf, accretes enough material to raise its core temperature enough to reignite nuclear fusion, causing a runaway reaction that leads to a supernova and 2) If a massive star (> 8 solar masses) undergoes core collapse, it will release a tremendous amount of potential gravitational energy, blowing the outer layers of the star into space at speeds approaching 10% the speed of light. A supernova can release as much energy in a month as our sun does in 10 billion years. Supernovae can outshine entire galaxies and may be visible to the naked eye for a period ranging from weeks to years.

Both novae and supernovae form shells of heated material that are ejected into space. Those formed from novae are much less massive and have a lifetime of a few hundred years. Supernovae remnants have much longer lifetimes and will expand into the surrounding medium for hundreds to thousands of years before slowing to local speeds. The interested reader is directed to Arnett (1996) for an in-depth discussion of supernovae physics.

Bright novae and supernovae would have had an impact on the people and cultures that witnessed them. Several have been visible to the naked eye over the last few thousand years, some of which appeared brighter than the brightest stars. Yet accounts of these events in oral (non-literate) cultures around the world remain scant in the literature.

## 3    HISTORICAL NOVAE AND SUPERNOVAE

Seven bright novae (magV = +2 or brighter) have been visible from Earth since 1890.[1] Only one – V603 Aquilae – rivalled Sirius (Alpha Canis Majoris, magV = –1.46) in brightness, reaching a visual magnitude (magV) of –1.4 in 1918 (Harrison et al., 2013). Nine supernovae are known have been visible to the naked eye in the last 2,000 years (Green, 2002), seven of which had an apparent magnitude of 0 or brighter (see Table 1).

Supernovae are generally named after the year in which they were visible (or believed to have been visible). For example, a supernova that appeared in the year 1054 CE is named 'SN 1054'. Recorded supernovae visible with the naked eye appear roughly once every 250 years on average. Historical supernovae have been recorded by literate cultures such as the Chinese, Koreans, and Romans (Shen, 1969; Chu, 1968; Stothers, 1977; Clark and Stephenson, 1982). A list of currently known bright supernovae recorded over the last 2,000 years, plus a few bright supernovae that are believed to have been visible before this time, is given in Table 1 in order of age (oldest to youngest).

*Table 1: Known bright (minimum magV 2.0) supernovae since 1 CE, plus five supernovae known to have been visible during human (pre)history. The peak magnitude of RX is not known, as its distance is controversial. The data are shown in order of age, citing their designation, age or year discovered (CE =Common Era, BP =Before Present), current home constellation, celestial coordinates (right ascension and declination in J2000 from SIMBAD: http://simbad.u-strasbg.fr), and the peak visual magnitude, if known (there is some uncertainty in these estimates – the values provided are estimates only). Data for SN 185 to SN 1604 (Stephenson and Green, 2002), RX J0852.0-4622 (Aschenbach et al., 1999), E0102-72 (Hughes et al., 2000), Puppis-A (Winkler et al., 1988), HBH9 (Leahy and Tian, 2007), Veil (Blair et al., 2009), and Vela (Cha et al., 1999). The magV of Puppis-A was calculated using current estimates of its distance[2].*

| Name | Location | RA | DEC | magV | D (ly) | Type | Age |
|---|---|---|---|---|---|---|---|
| Vela | Vela | 08h 34m | −45° 50′ | −10 ? | 815±98 | - | 11000-12300 |
| Veil | Cygnus | 20h 45m | +30° 42′ |  | 1470 | - | ~5000 |
| HBH 9 | Auriga | 05h 01m | +46° 40' |  | 2608±1304 |  | 4000-7000 |
| Puppis A | Puppis | 08h 24m | −42° 59′ | −1.4 ? | 7000 | - | ~3700 |
| E0102-72 | Tucana | 01h 04m | −72° 01′ | +2 | 190000 | - | 1000–2650 |
| RX J0852.0-4622 | Vela | 08h 52m | −46° 22′ |  | 700? | - | 680-1100 |
| SN 185 | Circinus | 14h 43m | −62° 30′ | <−8 ? | 8200 | Ia(?) | 185 CE |
| SN 386 | Sagittarius | 18h 14m | −19° 46′ | +1 | 14700 | II | 386 CE |
| SN 393 | Scorpius | 17h 14m | −39° 48′ | −1 | 34000 | - | 393 CE |
| SN 1006 | Lupus | 15h 02m | −42° 06′ | −7.5 | 7200 | Ia | 1006 CE |
| SN 1054 | Taurus | 05h 35m | +22° 01′ | −4 | 6500 | II | 1054 CE |
| SN 1181 | Cassiopeia | 02h 06m | +64° 50′ | −1 | 8500 | - | 1181 CE |
| SN 1572 | Cassiopeia | 00h 25m | +64° 09′ | −4 | 8000 | Ia | 1572 CE |
| SN 1604 | Ophiuchus | 17h 30m | −21° 29′ | −2.5 | 14000 | Ia | 1604 CE |

Most of these records were taken from cultures in the northern hemisphere. Since a significant portion of the galactic plane, as well as the Large and Small Magellanic Clouds, are only visible at more southerly latitudes, some bright supernovae were not recorded by literate cultures.

Since indigenous cultures around the world were keen observers of the night sky (e.g. Ruggles, 2011), we expect to find descriptions of novae/supernovae in indigenous oral or material culture. Many claims of these records have been made. The next section establishes criteria for confirming these accounts, followed by a brief survey of novae/supernovae claims from around the world. Finally, we analyse an oral tradition from Australia that describes the appearance of a bright new star in the sky.

# 4 IDENTIFYING NOVAE AND SUPERNOVAE IN INDIGENOUS CULTURAL TRADITIONS

Since transient celestial phenomena, such as comets, meteors, and eclipses, are recorded in oral and material traditions by cultures around the world, it is likely that supernovae were also recorded. But identifying a clear example is difficult, as two main problems impede our ability to directly associate an oral tradition or art form with a particular supernova event: (1) descriptions in oral traditions are generally ambiguous; and (2) motifs in material culture are sometimes open to interpretation. Because of this, misidentifications and misinterpretations of supernovae-in-culture are common. This necessitates a list of criteria required to confirm a novae/supernovae in oral traditions (O), material culture (M), or both (OM):

(1) O: There is a description of a "new star" appearing in the sky;
(2) OM: The location on Earth from which the "new star" was seen;
(3) OM: The period in time when the "new star" appeared;
(4) OM: The location of the "new star" in the sky.
(5) M: Evidence that the motif represents a star.
(6) OM: Novae/supernova remnant located where "new star" was visible.

In many Indigenous Australian cultures, as with other global cultures, spirits of the deceased take the form of stars (Johnson, 1998: 16-19). Therefore, stories that describe the appearance of a bright new star in the sky may be a reflection of this belief and not an account of a nova or supernova. Establishing a date of origin for an oral tradition is also problematic since many indigenous traditions are not set in linear time. Frequently, they refer to a creation period "long ago". For novae/supernovae representations in material culture, archaeometric-dating methods could potentially be used to identify a time period from which the account may have taken place, assuming the artefact or artwork was made when the supernova was visible. But without ethnographic or historical evidence, connecting material culture to these astronomical events is speculative. Regarding criterion #6, the supernova record is incomplete, so a cultural description of a supernova event could assist astronomers in locating a supernova remnant, particularly in the southern latitude skies where records are fewer than in the north. Novae remnants also have much shorter lives than supernovae remnants. This could make identifying a nova difficult (although necessary for confirmation). Although not a set criterion, it would be helpful (if applicable) that an oral tradition acknowledge that the "new star" is no longer visible to the naked eye.

Each criterion has a potential "gray area", such as the time the new star was visible (does the time estimated need to be within a day, a year, a century?) or the star's location in the sky (within a degree or two of the supernova remnant or just in that general part of the sky?). Even providing evidence that a motif represents a star can be challenging. These criteria serve only as a guide, and any supernova account that is weighed against these criteria will need to be assessed individually.

These criteria will now be applied to proposed supernovae accounts from around the world, followed by a survey of possible accounts from Australia.

## 5 PROPOSED NOVAE AND SUPERNOVAE ACCOUNTS IN CULTURE

### 5.1 BOLIVIA: VELA SN

George Michanowsky proposed that a supernova was witnessed by Indigenous Bolivians and recorded in their rock art (Anonymous, 1973). A petroglyph on a flat rock is believed to be associated with an annual festival. The petroglyph shows four small circles, flanked by two larger circles. Michanowsky believes the four smaller circles represent the stars of the "false cross" on the border of Carinae and Vela, and the two larger circles represent Canopus (Alpha Carinae) and the Vela supernova. He cites the name of this part of the sky (the region of the Gum Nebula) by local Indigenous people as *Lakha Manta*, apparently meaning *The Gateway to Hell*, and mentions some local astronomical traditions describing two celestial dogs that chased an ostrich across the sky, which they killed in the constellation Vela. Michanowsky's interpretation is speculative and there is no ethnographic evidence to support this hypothesis. The Indigenous people of the area claim no knowledge regarding the petroglyph's meaning or any connection to their astronomical traditions.

### 5.2 IRAQ (MESOPOTAMIA): VELA SN

George Michanowsky (1977) also claimed that the Vela supernova was witnessed by the Mesopotamians and recorded on their artefacts. He claims the Vela supernova was recorded as the *"giant star of the god Ea in the constellation of Vela of the god Ea"* on a clay tablet. Michanowsky's reasoning for interpreting this symbol (according to his translations) is that the "constellation to the north is the "Exalted Lady": the constellation Puppis. This would have put the "giant star" on the border of the constellations Vela and Puppis, where the Vela supernova remnant is located. Michanowsky cited an age of ~6,000 years for the Vela supernova, despite the age estimated at the time being between 6,000 and 11,000 years. Besides the number of assumptions and interpretations necessary for this hypothesis to be true, modern astrophysical observations show that the Vela supernova was visible in the sky > 10,000 years ago (Cha et al., 1999), not the lower limit of 6,000 years as cited by Michanowsky.[3]

### 5.3 BURZAHOM, INDIA: SN HBH 9

According to Iqbal et al. (2009) and Joglekar et al. (2011), Neolithic art at Burzahom (near Srinagar, India) depicts two celestial objects above a pair of hunters (Figure 1E). They cite the age of the art as 3,000-1,500 BCE and Joglekar et al. suggest that the motifs represent a full (or near full) moon and a supernova. Searching for supernovae remnants dating to within this time period and near the ecliptic, Joglekar et al. propose the candidate is supernova HBH 9, which dates to ~5,000-2,000 BCE (Leahy and Tian, 2007). The meaning of the rock art is open to interpretation and the hypothesis that this represents a supernova is based on loose assumptions.

### 5.4 NEW ZEALAND: SN 185

Green and Orchiston (2004) propose that *Mahutonga*, a star with no modern counterpart that is described in the astronomical traditions of the Maori of Aotearoa (New Zealand), may be a description of a supernova. Mahutonga means "a star of the south that remains invisible." The relationship between the term "Mahu" and the constellation Crux suggests it is related to this part of the sky. Green and Orchiston search for candidate supernovae near Crux that would have been visible over the last millennium when the Maori arrived to New Zealand. Due to its location near Alpha Centauri, the authors suggest SN 185 is the best candidate. This would mean that the Mahutonga reference is proto-Polynesian in origin, as this supernova was visible nearly 1,000 years before the Maori colonised New Zealand around 1280 CE (Lowe, 2008). It is also possible the historical supernovae record is incomplete and Mahutonga may refer to a currently unknown supernova in that region of the sky.

### 5.5 ARIZONA, USA: SN 1006

Barentine (2006) suggests that a Hohokam petroglyph outside of Phoenix, Arizona (Figure 1A) represents SN 1006. The area was occupied from 500-1100 CE and Barentine claims the event is depicted "in context with recognizable asterisms," such as Scorpius. The claim was challenged by archaeoastronomers Anthony Aveni and Edwin Krupp as a misinterpretation of a common motif (Cull, 2006). This reference is a conference abstract and nothing else has been published in the literature. Until further evidence is presented, this interpretation is speculative.

### 5.6 GUAM: SN 1054

Villaverde (2000a) claims that an oral tradition from the Micronesian island of Guam may describe the appearance of a new star in the sky that rivalled Venus in brightness. Villaverde believes this new star is SN 1054. Researcher Frank Guerrero claims that pictograms at Ritidian Cave on the northern end of Guam (Figure 1B) represent events that occurred in the constellation Auriga, which is adjacent to Taurus where SN 1054 was visible (Villaverde, 2000b). The cave contains pictograms showing a lunar calendar estimated to have been made some 4,000 years ago (Iping, 1999). This research is still in progress and has not yet been subject to peer-review.

### 5.7 NEW MEXICO, USA: SN 1054

The most famous – and controversial – claim of a supernova being represented in material culture is that of Anasazi rock art in Chaco Canyon, New Mexico (e.g. Brandt and Williamson, 1979; Brecher *et al.*, 1983) (Figure 1C). Some researchers claim that a motif near a hand stencil and a crescent moon represents SN 1054, which was visible in the sky from the site during the period of Anasazi culture and would have appeared below the crescent moon on the date it was first visible, as noted by Chinese historical records. Despite being wildly popular with the public, many researchers within the anthropological and archaeological communities do not support this interpretation, refuting these claims as a misrepresentation of the crescent Moon and Venus (e.g. Koenig, 1979; Collins et al., 1999; and Schaefer, 2006, among many).

### 5.8 SRINAGAR, INDIA: SN 1604

Sule et al. (2011) claim to have identified SN 1604 in a mural on a doorway arch (Figure 1D) in the mosque of Madani, in Srinagar, India. According Sule et al., the mural shows the alleged supernova as a dragon's head on the tail of the constellation Sagittarius. This interpretation has been challenged as a misidentification of Islamic astrological iconography (van Gent, 2013).

### 5.9 VICTORIA, AUSTRALIA: 1843 ERUPTION OF ETA CARINAE

In the 1840s, the hyper-massive luminous blue variable (LBV) star Eta Carinae went through a period of instability and ejected 20 solar masses of stellar material into space, brightening to such an extent that it appeared as the second brightest star in the night sky before fading to invisibility (Smith and Frew, 2011). Hamacher and Frew (2010) demonstrated that the event was incorporated into the oral traditions of the Boorong Aboriginal people in northwestern Victoria, near Lake Tyrrell. This account was confirmed because an Englishman named William E. Stanbridge recorded a detailed description about Boorong astronomy from two Boorong men in the 1840s. The description includes the brightness, colour, position, and catalogue number of the star (Stanbridge, 1858, 1861). It should be noted that this star <u>does not classify</u> as either a nova or a supernova. It was an eruptive variable that did not destroy the star. This is commonly called a "supernova impostor" event.

### 5.10 AUSTRALIA: UNIDENTIFIED

Murdin (1981) speculates that a circular "bicycle wheel" motif in Aboriginal rock art in Sturts Meadows, western New South Wales, Australia, represents a supernova (Figure 1F). These motifs, as well as others sometimes called "sunburst" designs, are found in rock art across Australia (e.g. McCarthy, 1983: 163; Flood, 1997: 183, 185, 208, 241). Two Aboriginal petroglyphs in Lane Cove National Park in the northern suburbs of Sydney (known to the author) show a crescent and a "sunburst" design, reminiscent of the Chaco Canyon rock art. The meaning of these "sunburst" motifs varies, and in many places (such as Lane Cove) their meaning is unknown. None are known to depict supernovae.

### 5.11 SUMMARY

Only one of the accounts described in this section confirms the sudden brightening of a star in oral traditions: the "supernova-impostor" eruption of Eta Carinae in 1843. This event was witnessed by the Boorong people of western Victoria, Australia and incorporated into their oral traditions. This is confirmed because Stanbridge recorded the required data, such as the star's physical appearance, position in the sky, and catalogue number, during the star's outburst. However, the Great Eruption of Eta Carinae was not a proper nova or supernova. All other cases described above are inconclusive and require further evidence before they could be confirmed as nova/supernova accounts.

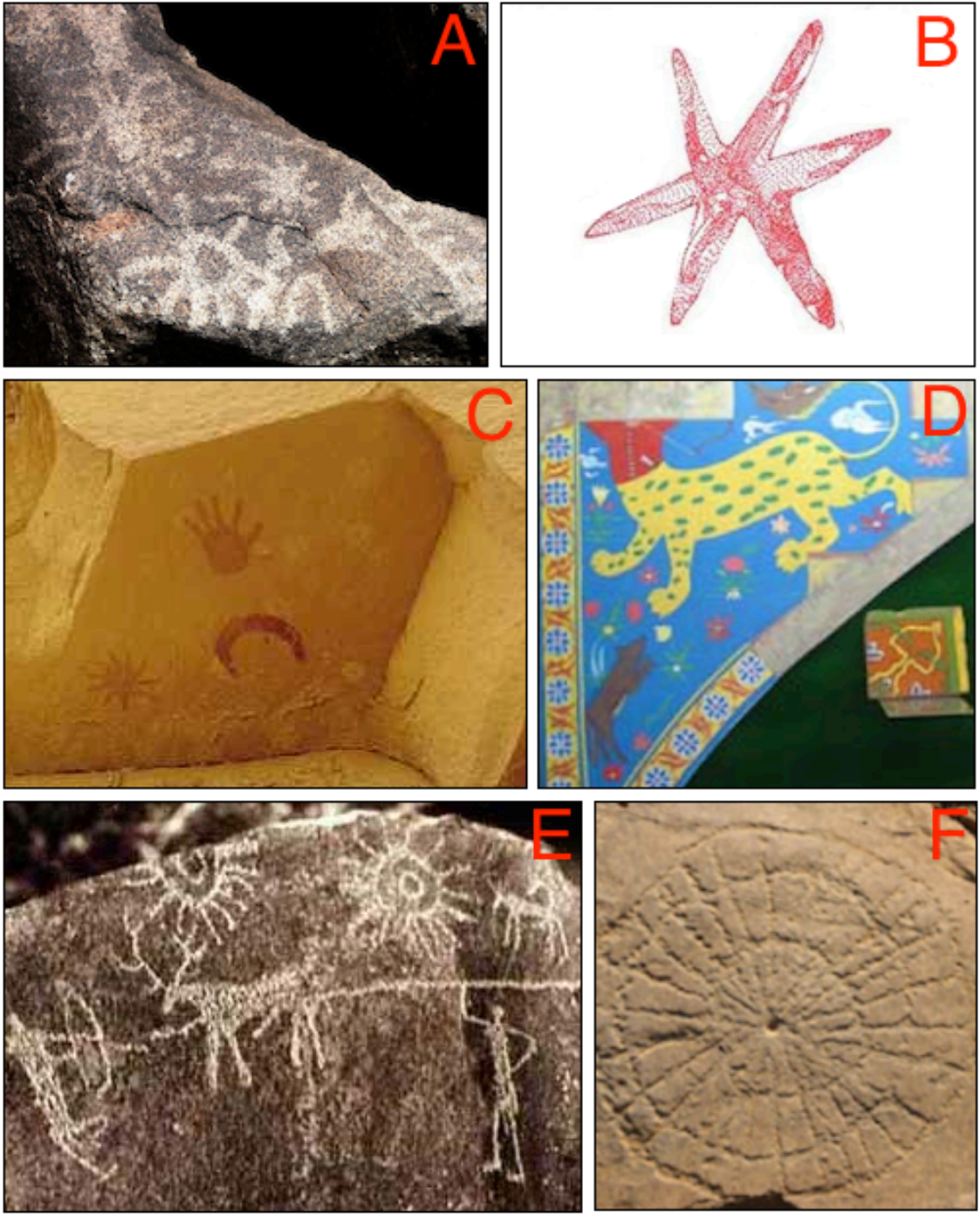

*Figure 1: **A**: Hohokam rock art, Arizona, USA (photo: J. Barentine), **B**: Ritidian Cave rock art, Guam (illustration by Michael S. Argenal), **C**: Chaco Canyon rock art, New Mexico, USA (photo: R. Lussier). **D**: Mural in the Mosque of Madani, Srinagar, India (after Sule et al., 2011), **E**: Neolithic rock art, Burzahom,*

*India (photo: http://ignca.nic.in).* ***F****: Aboriginal rock art, New South Wales, Australia (after Murdin, 1981: 477).*

The following section analyses a possible supernova account from Australia. As with the accounts in this section, the story is analysed to demonstrate the challenges and difficulties in linking oral traditions to historical supernovae.

## 6     A SUPERNOVA ACCOUNT FROM AUSTRALIA?

Literature regarding Australian Indigenous oral traditions was surveyed for references to possible supernovae. One story was identified that indicated the appearance of a bright new star in the sky. The story, entitled "The Fisherman Brothers", is from the Yolngu people of Arnhem Land in the Northern Territory of Australia (Wells, 1973: 31-36). In the story, two brothers, Nuruguya-mirri (the older) and Napiranbiru (the younger), were fishing in a canoe when a storm hit, wrecking their boat. The Nuruguya-mirri was stronger and helped Napiranbiru reach the shore, but drowned while saving him. To honour Nuruguya-mirri's courage, the community held a ceremony. That night, the people saw "a bright star shining, new and sparkling" in the sky. The people said the new star was Nuruguya-mirri. When Napiranbiru was an old man and passed away, he asked the spirits to place him in the sky with his brother. The brothers are now two bright stars close together on the bank of the sky-river, called *Milnguya* (Milky Way). A nearby faint star is the brothers' little campfire. These stars have been shining in the sky "for longer than anyone can remember," (p. 31).

Variations of the story are found across Arnhem Land. Mountford (1956: 496-499) cites a similar identical story from Blue Mud Bay, Arnhem Land. In this account, two brothers are fishing in a bark canoe near Woodah Island, northwest of Groote Eylandt off the southeastern coast of Arnhem Land. On their way home, a storm hit and the waves capsized the brothers' canoe. In this account, the younger brother (*Jikawana*) dies first, followed by the older brother (*Nungumiri*). The story of the brothers' demise is reflected in the sky. The brothers, the fish, and their canoe are seen as dark absorption nebulae ("starless spaces") in the Milky Way. According to Mountford (*ibid*: 498), the younger brother is a dark nebula in the Milky Way within the constellation Serpens, and the elder brother is a dark nebula near Sagittarius. The rock on which the elder brother collapsed is seen as a portion of the Milky Way near Theta Serpens, and their canoe is represented by a line of four small stars near Antares (Alpha Scorpii). This story is represented in Aboriginal bark paintings from Arnhem Land (see Figure 2).

Additional sources were examined for variations of this story (e.g. Allen, 1976; Warner, 1937; Mountford, 1956; Chaseling, 1957; Harney and Elkin, 1968; Berndt and Berndt, 1989). Wells' account is the only version that mentions the appearance of a bright new star, although Allen mentions a bright flash of light in the sky. It is unclear if this account was recorded exactly as the Aboriginal custodians told it or if Wells used a degree of "poetic license" in retelling the story, inserting the description of a "new star" herself.

According to Wells, the bright new star is still visible as one of two stars close together along the sky river (Milnguya), with a fainter star nearby representing the brother's celestial campfire (Wells, 1973: 31). The identity of the two stars representing the brothers is not given, but is most likely the optical double stars Shaula and Lesath (λ and υ Scorpii, magV = +1.6 and +2.7, respectively), which are both commonplace in Yolngu and other Aboriginal astronomical traditions (Mountford, 1956: 481, 501; Mountford, 1976: 459-460; Stanbridge, 1858: 139; Johnson, 1998: 71, 74; Fredrick, 2008: 118). Shaula is the second brightest star in Scorpius and Lesath is separated from Shaula by ~ 0.6°. They both lie within the galactic bulge near the border of a dark dust lane. To the Aboriginal people of Groote Eylandt, the acronychal rising of Shaula and Lesath (at the end of April) signals the end of the wet season and the start of *mamariga* – a dry season marked by southeasterly winds (Mountford, 1956: 481). These stars are seen as the children of the planets Venus (*Barnimbida*, father) and Jupiter (*Duwardwara*, mother). Mountford does not specify that these are the boys in the Fishermen Brothers story, but variations of stories, or stories of the same stars, exist within the same language groups. If the identity of the two brothers is Shaula and Lesath, the faint star (the brothers' little campfire in the sky) could be HR 6583 (Spectral Type K0/K1III, magV = +5.53), Q Scorpii (G8/K0III, +4.27), or HR 6501 (K0III, +5.99).

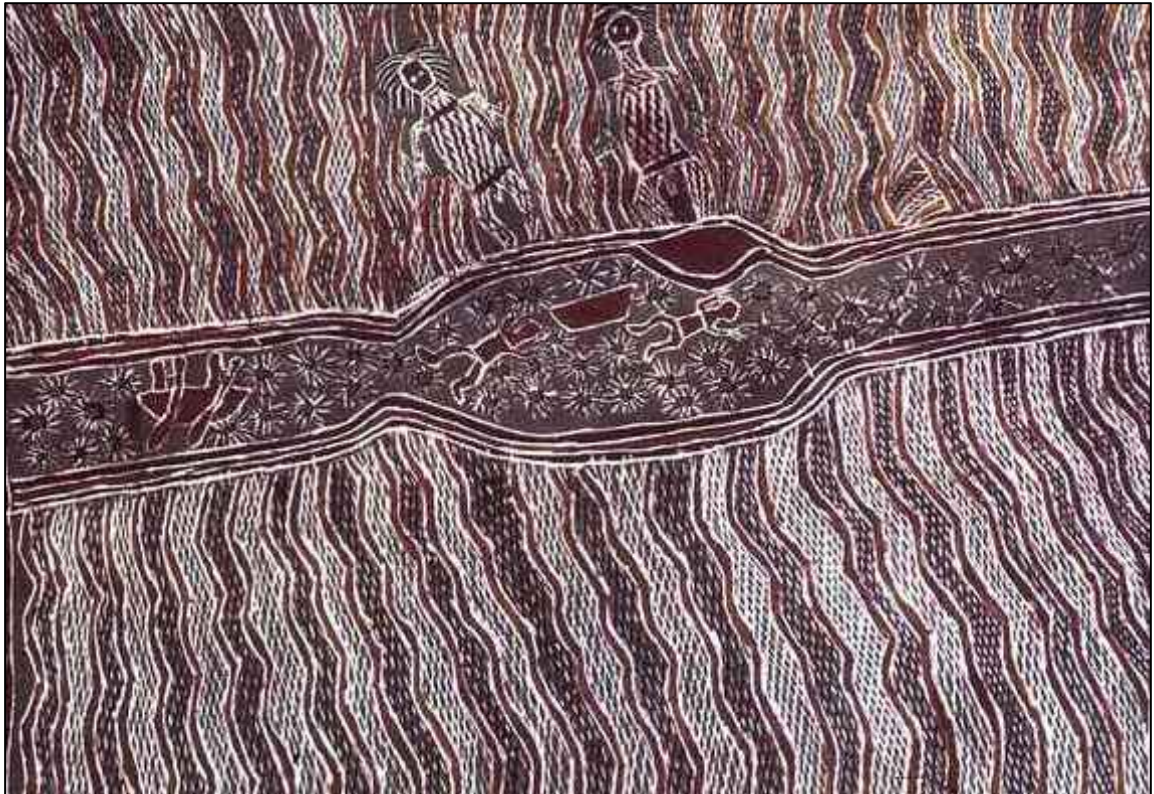

*Figure 2: A Yolngu bark painting of the Fisherman Brothers in the Milky Way. After Mountford (1956: 499), Plate 157-A (cropped). See also Plate 157-C.*

The story indicates that all three stars are still visible today. The criteria set above are not met, meaning this is an unconfirmed supernova claim. It is worth noting that a bright supernova was visible within 4° of Shaula and Lesath in the year 393 CE. SN 393 occurred in a dark patch within the tail of Scorpius (Figure 3). SN 393 was recorded by Chinese astronomers in the asterism of Wei (the tail of Scorpius) in the second lunar month (between 27 February and 28 March) and was reportedly visible for about seven months before fading from visibility (Clark and Stephenson, 1975; Wang, 2006). It appeared as a bright star with an estimated magV of −1 at its peak, making it the second brightest star in the sky after Sirius before fading from visibility. It is unknown if this supernova relates to the Fisherman Brothers story, but the connection seems improbable.

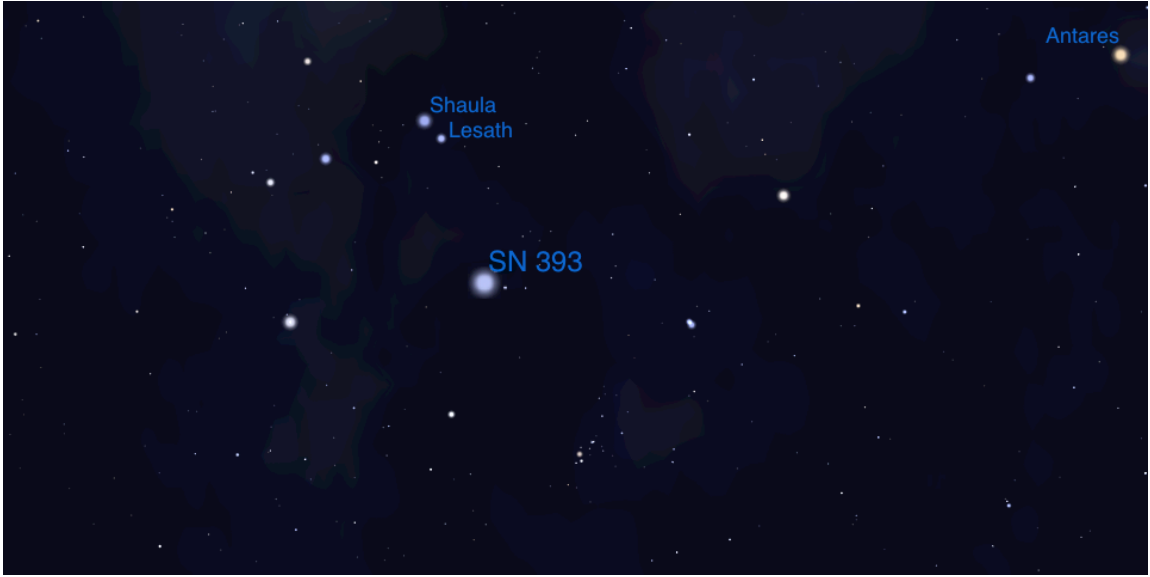

*Figure 3: SN 393 as seen in the tail of Scorpius near the stars Shaula and Lesath. Image from Stellarium with labels included by the author.*

## 7  DISCUSSION AND CONCLUDING REMARKS

We are certain that ancient and Indigenous people witnessed novae and supernovae, and we strongly believe that these events were incorporated into their oral traditions and possibly material culture. There is only solid evidence that the Boorong people of western Victoria noted the Eta Carinae supernova-impostor event in the 1840s and incorporated it into their astronomical traditions (Hamacher and Frew, 2010). Without the written information provided by W.E. Stanbridge, confirming this record would be impossible. None of the remaining claims presented in this paper are confirmed and many are either not accepted by the academic community or have not yet been subject to peer review (e.g. Barentine, 2006; Michanowsky, 1973, 1977; Villaverde, 2000a, 2000b).

The Fishermen Brothers story from Australia indicates the appearance of a bright new star in the sky. The location from which the story was taken is known, but the location of the "new star" in the sky is inferred. The time period in which the star appeared is unknown, and only one story describes the appearance of a bright new star. From the criteria set in Section 4, it is considered unconfirmed.

This paper establishes criteria necessary to identify novae/supernovae in oral or material culture. It should be emphasised that attempts to link oral traditions or material culture with novae/supernovae are worthwhile. The identification of these phenomena in oral or material culture would benefit our understanding of cultural astronomy and Indigenous Knowledge traditions. These accounts could also potentially lead astronomers to unrecorded supernovae remnants.

## 8   ACKNOWLEDGEMENTS

The author would like to acknowledge the Indigenous elders and custodians of Australia. Special thanks to Steve Hutcheon and John McKim Malville for their valuable feedback and constructive criticism. This research made use of the TROVE database (trove.nla.gov.au), JSTOR (www.jstor.org), NASA Astrophysics Data System (adsabs.harvard.edu), and the Stellarium astronomical software package (www.stellarium.org).

## 9   NOTES

1. IAU Central Bureau for Astronomical Telegrams, CBAT List of Novae: http://www.cbat.eps.harvard.edu/nova_list.html

2. Using the standard equation m $= -5 + 5*\log_{10}(d) + M + A$, where m is the star's apparent magnitude, d is the star's distance in parsecs (2,147 pc), M is the star's absolute magnitude ($-17$ for a Type II SN), and A is galactic extinction (1.8 magnitudes per kpc for stars in the galactic plane within a few kpc of Earth). This only serves as a rough estimate.

3. If Michanowsky's translation of the position of the "giant star" is accurate, and assuming the "giant star" was a supernova, another possibility could be the Puppis-A supernova, which was visible in the same part of the sky ~3,700 years ago. At 8.5 times the distance of the Vela remnant, Puppis-A would have been much fainter than Vela, but still bright (magV $= -1.4$, approximately as bright as Sirius). It should be noted that Michanowsky's claims are not widely accepted and have been criticized by the academic community (e.g. Huyghe, 1981).

## ABOUT THER AUTHOR

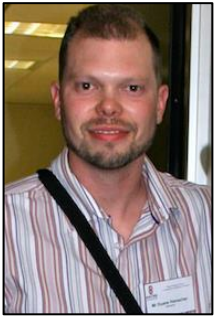

Dr Duane Hamacher is a Lecturer and ARC Discovery Early Career Researcher in the Nura Gili Indigenous Programs Unit at the University of New South Wales in Sydney, Australia. His research and teaching focuses on cultural astronomy and geomythology, with a specialization on Indigenous Australia and Oceania. He earned graduate degrees in astrophysics and Indigenous studies and works as an astronomy educator and consultant curator at Sydney Observatory.